\documentclass{aastex}
\usepackage{emulateapj5,times,mathptm}
\journalinfo{The Astronomical Journal, 2001 November, astro-ph/0108001}
\slugcomment{Received 2001 June 11; accepted 2001 July 30}

\shorttitle{{\it CHANDRA} OBSERVATIONS OF $Z>4$ QUASARS}
\shortauthors{VIGNALI ET AL.}

\newcommand{\ltsima}{$\; \buildrel < \over \sim \;$}
\newcommand{\simlt}{\lower.5ex\hbox{\ltsima}}
\newcommand{\gtsima}{$\; \buildrel > \over \sim \;$}
\newcommand{\simgt}{\lower.5ex\hbox{\gtsima}}

\newcommand{\ROSAT}{{\it ROSAT} }

\newcommand{\ASCA}{{\it ASCA} }
\newcommand{\CHANDRA}{{\it Chandra} }
\newcommand{\XMM}{{\it XMM-Newton} }
\newcommand{\XEUS}{{\it XEUS} }
\newcommand{\CONS}{{\it Constellation-X} }

\begin{document}

\title{Exploratory Chandra Observations of the Highest-Redshift Quasars: 
X-rays from the Dawn of the Modern Universe}

\author{
C. Vignali,\altaffilmark{1} 
W.~N. Brandt,\altaffilmark{1} 
X. Fan,\altaffilmark{2} 
J.~E. Gunn,\altaffilmark{3} 
S. Kaspi,\altaffilmark{1} 
D.~P. Schneider,\altaffilmark{1} 
and \hbox{Michael~A. Strauss\altaffilmark{3}}
}

\altaffiltext{1}{Department of Astronomy and Astrophysics, The Pennsylvania State University, 
525 Davey Laboratory, University Park, PA 16802 \\
({\tt chris@astro.psu.edu, niel@astro.psu.edu, shai@astro.psu.edu, and dps@astro.psu.edu}).}
\altaffiltext{2}{Institute for Advanced Study, Olden Lane, Princeton, NJ 08540-0631 
({\tt fan@sns.ias.edu}).}
\altaffiltext{3}{Princeton University Observatory, Peyton Hall, Princeton, NJ 08544-1001 \\
({\tt jeg@astro.princeton.edu and strauss@astro.princeton.edu}).}


\begin{abstract}

We report exploratory {\it Chandra} observations of 14 high-redshift \hbox{($z$=4.06--5.27)}, 
optically selected quasars. Ten of these quasars are detected, increasing the number 
of \hbox{$z>4$} X-ray detected quasars by 71\%. 
Our detections include four of the five highest-redshift X-ray detected quasars to date, among them 
SDSSp~J021043.17$-$001818.4, the highest-redshift ($z=4.77$) radio-loud quasar detected in the X-ray band. 
The four undetected objects are 
the Broad Absorption Line quasars SDSSp~J112956.10$-$014212.4 and SDSSp~160501.21$-$011220.0,  
the weak emission-line quasar SDSSp~J153259.96$-$003944.1, and the quasar PSS~1435$+$3057. 
A comparison of the quasars' spectral energy distributions (by means of 
the optical-to-X-ray spectral index, $\alpha_{\rm ox}$) with those of 
\hbox{lower-redshift} samples indicates that the \CHANDRA quasars are X-ray 
fainter by a factor of $\approx2$. \hbox{X-ray} faintness could be associated with the presence of large 
amounts of gas in the primeval galaxies harboring these high-redshift quasars, as 
suggested by recent studies conducted on 
$z>4$ quasars in other bands. 
Using the current \CHANDRA data, predictions for the next generation of X-ray observatories, 
\hbox{\CONS} and {\it XEUS}, are also provided. 
\end{abstract}


\section{Introduction}

The discovery of quasars at $z>4$, first accomplished by Warren et al. (1987), 
has recently been the subject of particular interest 
because of the first results coming from the Sloan Digital Sky Survey 
(SDSS; York et~al. 2000). In the last few years, the number of known 
$z>4.6$ quasars has more than tripled (e.g., Fan et~al. 1999a, 2000a; 
Anderson et~al. 2001). About 1000 \hbox{$z>4$} SDSS quasars are expected over the 
next five years, $\approx 100$ of which should be at $z>5$ (Fan 1999; Schneider 1999).

Studies of quasars at the highest redshifts are important for addressing
a number of key astrophysical and cosmological issues. 
High-redshift quasar studies allow investigation of the 
formation of galaxies and large-scale structures at epochs when 
the Universe was only \hbox{5--10\%} of its present age (e.g., Efstathiou \& Rees 1988; 
Turner 1991; Small \& Blandford 1992; Haehnelt \& Rees 1993). From the Eddington 
limit, luminous quasars are expected to host \hbox{$\ga10^{8-9}$ M$_{\odot}$} black 
holes. Such objects are probably associated with rare high peaks in the primordial 
density field, and these peaks should be strongly clustered (e.g., Kaiser 1984). 
The strong evolution of high-redshift quasars can therefore provide clues about 
the processes by which the remarkably homogeneous $z\approx1400$ Universe revealed 
by the cosmic microwave background is transformed into the inhomogeneous Universe seen today. 
Furthermore, the emission-line strengths of high-redshift quasars 
can be used as diagnostics of gas metallicity. If the gas in high-redshift quasars is 
related to the interstellar material of their young host galaxies, the quasars can be 
used to constrain the star-formation history and chemical enrichment processes 
of primordial galactic environments (e.g., Hamann \& Ferland 1999; 
Dietrich \& Hamann 2001). 

To date, quasars at $z>4$ have been extensively studied at optical 
wavelengths (e.g., Schneider, Schmidt, \& Gunn 1989, 1991; Kennefick et~al. 1995; 
Kennefick, Djorgovski, \& Meylan 1996; Storrie-Lombardi et~al. 1996, 2001; Fan et al. 2001), in the radio band 
(e.g., Schmidt et~al. 1995; Hook, McMahon, \& Shaver 1999; Stern et~al. 2000; Yun et~al. 2000) and, 
to a lesser extent, in the sub-mm and mm range (e.g., McMahon et~al. 1994, 1999; 
Omont et~al. 1996a, 2001; Guilloteau et~al. 1999; Benford et~al. 1999; Carilli et al. 2001). 
In contrast, our knowledge of the X-ray properties of quasars at $z>4$ is 
quite limited. At present, only 14 quasars and one Seyfert galaxy 
at $z>4$ have published X-ray detections; three of them are X-ray selected. 
It has only been possible to obtain X-ray spectra 
for four $z>4$ quasars; all four are blazars whose X-ray spectral 
and variability properties are probably jet-induced (e.g., Moran \& Helfand 1997; Yuan et~al. 2000; 
Fabian et~al. 2001a,b). The most systematic X-ray study of $z>4$ 
quasars has been carried out by Kaspi, Brandt, \& Schneider (2000; hereafter KBS) 
using archival \ROSAT data. While the most basic X-ray properties of $z>4$ quasars 
(e.g., $\alpha_{\rm ox}$, the slope of a nominal power law connecting 2500~\AA\ and 
2~keV in the rest frame) appear to be consistent with those of lower-redshift samples, 
the constraints on their X-ray spectral shapes are still weak. 

X-ray studies of the highest-redshift quasars can probe both 
their central power sources and their environments. 
Measurement of the intrinsic \hbox{X-ray} continuum's shape, normalization relative to 
the rest of the spectral energy distribution (SED), and variability can provide
information on the inner accretion disk and its corona, and thus ultimately
about how the black hole is being fed. There are already some
reports that the \hbox{X-ray} continuum shapes of quasars may evolve with redshift
at $z\simlt 2.5$ (e.g., Vignali et~al. 1999; Blair et~al. 2000); these 
intriguing results require further study and extension to the highest 
redshifts. 
\begin{table*}[t]
\footnotesize
\caption{{\it CHANDRA} Observation Log\label{tab1}}
\begin{center}
\begin{tabular}{lccccccc}
\hline
\hline
Object & & Optical & Optical & $\Delta_{\rm Opt-X}$ & Obs. & Exp. Time & \\
Name & $z$ & $\alpha_{2000}$ & $\delta_{2000}$ & (arcsec) & Date & (ks) & Ref. \\
\hline
SDSSp~J021043.17$-$001818.4 & 4.77 & 02 10 43.2 & $-$00 18 18.4 & 0.6 & 2001/01/09 & 
4.95 & (1) \\
SDSSp~J021102.72$-$000910.3 & 4.90 & 02 11 02.7 & $-$00 09 10.3 & 0.3 & 2001/01/09 & 
4.95 & (2) \\
BRI~0241$-$0146\tablenotemark{a} & 4.06 & 02 44 01.9 & $-$01 34 04.2 & 0.2 & 2000/03/11 & 7.37 & (3) \\
PSS~0248$+$1802\tablenotemark{a} & 4.43 & 02 48 54.3 & $+$18 02 49.2 & 0.7 & 1999/12/27 & 1.73 & (4) \\
BRI~1033$-$0327\tablenotemark{a} & 4.51 & 10 36 23.8 & $-$03 43 19.3 & 0.9 & 2000/01/26 & 3.45 & (3) \\
PSS~1057$+$4555\tablenotemark{a} & 4.10 & 10 57 56.2 & $+$45 55 52.6 & 0.8 & 2000/06/14--15 & 2.81 & (5) \\
SDSSp~J112956.10$-$014212.4 & 4.85 & 11 29 56.1 & $-$01 42 12.4 & \nodata & 2001/03/26 & 4.88 & (6) \\
SDSSp~J120823.82$+$001027.7 & 5.27 & 12 08 23.8 & $+$00 10 27.7 & 0.6 & 2001/03/18 & 4.68 & (6) \\
PC~1247$+$3406 & 4.90 & 12 49 42.2 & $+$33 49 53.9 & 0.9 & 2001/03/24 & 4.68 & (7) \\
PSS~1317$+$3531\tablenotemark{a} & 4.36 & 13 17 43.2 & $+$35 31 32.0 & 0.3 & 2000/06/14 & 3.97 & (5) \\
PSS~1435$+$3057\tablenotemark{a} & 4.35 & 14 35 23.4 & $+$30 57 22.3 & \nodata & 2000/05/21--22 & 2.81 & (4) \\
PSS~1443$+$2724\tablenotemark{a} & 4.42 & 14 43 31.2 & $+$27 24 37.0 & 0.5 & 2000/06/12 & 2.17 & (4) \\
SDSSp~J153259.96$-$003944.1 & 4.62 & 15 33 00.0 & $-$00 39 44.1 & \nodata & 2001/03/26--27 & 5.15 & (8--9) \\
SDSSp~J160501.21$-$011220.0 & 4.92 & 16 05 01.2 & $-$01 12 20.0 & \nodata & 2001/06/24 & 4.64 & (9) \\
\hline
\end{tabular}
\vskip 2pt
\parbox{5.5in}
{\small\baselineskip 9pt
\footnotesize
\indent
{\sc Note. ---} 
The optical positions of the quasars not found by the SDSS have been 
taken from the Digital Sky Survey (DSS2). \\
$\rm ^a$ \CHANDRA archival observation. \\
{\sc References. ---}
(1) Fan et al. 2001; (2) Fan et al. 1999a; (3) Storrie-Lombardi et al. 1996; (4) Kennefick et al. 1995; 
(5) Kennefick, Djorgovski, \& de Carvalho 1995; 
(6) Zheng et al. 2000; (7) Schneider, Schmidt, \& Gunn 1991; (8) Fan et al. 1999b; (9) Fan et al. 2000a.
}
\end{center}
\vglue-0.7cm
\end{table*}
\normalsize
It is plausible that the X-ray continua of quasars may change at 
the highest redshifts. If our basic model for quasars is correct, then to be 
so luminous at such a young age, the \hbox{highest-redshift} quasars may be accreting matter near the 
Eddington limit where disk instabilities, ``trapping-radius'' effects, and 
other energetically important phenomena can arise. 
Regarding \hbox{high-redshift} quasar environments, 
there is growing evidence that the fraction of radio-loud 
quasars (RLQs) characterized by strong absorption in the X-ray band, 
increases with redshift (e.g., Elvis et~al. 1994; Fiore et~al. 1998; 
Reeves \& Turner 2000). The absorbing gas is thought to be physically 
associated with the quasars' environments, perhaps located in the host 
galaxy or entrained by the radio jets. The situation is far less clear 
for the radio-quiet quasars (RQQs), which show less of an absorption 
increase with redshift than do the RLQs (Fiore et~al. 1998). However, 
the constraints on any RQQ absorption-redshift connection are still 
quite loose and require substantial improvement, especially after the 
apparent discovery of significant X-ray absorption in some \hbox{$z\approx2$ RQQs} 
(Reeves \& Turner 2000). 

Motivated by these considerations, 
we have started a project to measure the X-ray properties of the most 
distant quasars, primarily those found by the SDSS, using the new generation 
of \hbox{X-ray} observatories. The SDSS multicolor selection method provides 
a sample of quasars that have been consistently selected in a well-defined manner. 
The most distant quasar published to date, 
\hbox{SDSSp~J104433.04--012502.2} at $z=5.80$ (Fan et~al. 2000b), has been successfully 
detected by \XMM (Brandt et~al. 2001a; hereafter B01) with a surprisingly low X-ray flux. 
The main goals of this work 
are to investigate whether the highest-redshift quasars show different X-ray properties 
than do local quasars, and to put the results into a comprehensive picture which 
links these objects to the growth of massive black holes from the 
first collapsed structures at the end of the cosmic ``dark age.''
The {\it Chandra X-ray Observatory} is ideal for this program, 
given its excellent angular resolution coupled with 
an extremely low background; this combination efficiently allows detection of very faint point sources 
with relatively low exposure times. 

Throughout this paper we adopt an $H_{0}$=70 km s$^{-1}$ Mpc$^{-1}$, $q_{0}$=0.5 and $\Lambda$=0 
cosmology. 

\section{Observations and data analysis}

\subsection{Chandra observations}

The present sample of high-redshift ($z>4$), optically selected quasars 
observed by \CHANDRA consists of 14 objects: seven were observed during 
Cycle~2, while for the remaining seven objects archival data have been analyzed. 
The observed sample contains many of the \hbox{$z>4.8$} quasars known at the time of the \CHANDRA Cycle~2 proposal round; 
the accepted Cycle~2 targets included all but one of the known luminous quasars at $z>4.8$ without previous 
or scheduled \XMM coverage (as well as one unusual \hbox{$z=4.62$} quasar 
lacking optical emission lines). 
The sources, along with their redshifts, optical positions, distances between the optical and 
X-ray positions, observation dates, and dead-time corrected exposure times, are presented in Table~1. 
In the following text the sources will be referred to through their abbreviated names. 
The 14 quasars were observed in 13 pointings; 
SDSS~0210$-$0018 was serendipitously observed in the field 
of the quasar SDSS~0211$-$0009 at an off-axis angle of $\approx 10.3$$\arcmin$.

\subsection{Chandra data analysis}

All of the sources were observed with the \CHANDRA Advanced 
CCD Imaging Spectrometer (ACIS; G.~P. Garmire et al., in preparation). 
ACIS consists of ten CCDs designed for efficient
X-ray detection and spectroscopy. Four of the CCDs 
(\hbox{ACIS-I}; CCDs I0--I3) are arranged in a 
$2\times 2$ array with each CCD tipped slightly to approximate the 
curved focal surface of the \CHANDRA High Resolution Mirror Assembly (HRMA). 
The remaining six CCDs (ACIS-S; CCDs S0--S5) are set in a linear array 
and tipped to approximate the Rowland circle of the objective gratings that 
can be inserted behind the HRMA. The CCD which lies on-axis in ACIS-S (S3) 
is orthogonal to the HRMA optical axis. It is a back-illuminated 
CCD with enhanced sensitivity in the soft X-ray band, and all but one source 
(the serendipitous one, SDSS~0210$-$0018, which was detected by the I2 front-illuminated CCD) 
were observed at its aimpoint. 
\begin{table*}[t]
\footnotesize
\caption{X-ray Counts in Four Energy Bands\label{tab2}}
\begin{center}
\begin{tabular}{lcccc}
\hline
\hline
 & \multicolumn{4}{c}{X-ray Counts} \\
\cline{2-5} \\
Object & [0.3--0.5 keV] & [0.5--2 keV] & [2--8 keV] & [0.5--8 keV] \\
\hline
SDSS~0210$-$0018 & $<3.0$ & 28.0$^{+7.0}_{-5.5}$ & {\phn}7.6$^{+4.5}_{-3.0}$ & 35.8$^{+7.8}_{-6.5}$ \\
SDSS~0211$-$0009 & $<3.0$ & {\phn}5.0$^{+3.4}_{-2.2}$ & $<3.0$ & {\phn}5.0$^{+3.4}_{-2.2}$ \\
BRI~0241$-$0146 & $<6.4$ & {\phn}9.0$^{+4.1}_{-2.9}$ & $<7.9$ & 11.9$^{+4.6}_{-3.4}$ \\
PSS~0248$+$1802 & 4.0$^{+3.2}_{-1.9}$ & {\phn}9.0$^{+4.1}_{-2.9}$ & {\phn}6.0$^{+3.6}_{-2.4}$ & 15.0$^{+5.0}_{-3.8}$ \\
BRI~1033$-$0327 & 5.0$^{+3.4}_{-2.2}$ & {\phn}8.0$^{+4.0}_{-2.8}$ & $<6.4$ & {\phn}9.9$^{+4.3}_{-3.1}$ \\
PSS~1057$+$4555 & 6.0$^{+3.6}_{-2.4}$ & 21.0$^{+5.7}_{-4.5}$ & {\phn}6.0$^{+3.6}_{-2.4}$ & 26.9$^{+6.3}_{-5.2}$ \\
SDSS~1129$-$0142 & $<3.0$ & $<3.0$ & $<3.0$ & $<3.0$ \\
SDSS~1208$+$0010 & $<3.0$ & $<4.8$ & $<4.8$ & {\phn}2.0$^{+2.6}_{-1.3}$ \\
PC~1247$+$3406 & $<4.8$ & 14.0$^{+4.8}_{-3.7}$ & {\phn}3.0$^{+2.9}_{-1.6}$ & 18.0$^{+5.3}_{-4.2}$ \\
PSS~1317$+$3531 & $<4.8$ & {\phn}3.0$^{+2.9}_{-1.6}$ & $<3.0$ & {\phn}3.9$^{+3.1}_{-1.9}$ \\
PSS~1435$+$3057 & $<3.0$ & $<3.0$ & $<3.0$ & $<3.0$ \\
PSS~1443$+$2724 & $<3.0$ & {\phn}7.0$^{+3.8}_{-3.6}$ & $<7.9$ & 10.0$^{+4.3}_{-3.1}$ \\
SDSS~1532$-$0039 & $<3.0$ & $<3.0$ & $<6.4$ & $<6.4$ \\
SDSS~1605$-$0112 & $<3.0$ & $<3.0$ & $<3.0$ & $<3.0$ \\
\hline
\end{tabular}
\vskip 2pt
\parbox{3.7in}
{\small\baselineskip 9pt
\footnotesize
\indent
{\sc Note. ---} 
Errors on the X-ray counts were computed according to Gehrels (1986). 
The upper limits are at the 95\% confidence level and were computed 
according to Kraft et al. (1991).\\
}
\end{center}
\vglue-0.7cm
\end{table*}
\normalsize
Each CCD subtends \hbox{an $8.3^{\prime}\times 8.3^{\prime}$} square on 
the sky, and the individual pixels of the CCDs subtend 
$\approx 0.5^{\prime\prime}\times 0.5^{\prime\prime} $ on the sky. 
The on-axis image quality of the telescope is FWHM $\approx0.5^{\prime\prime}$; 
this quantity increases to~$\approx 1.0^{\prime\prime}$ (critical sampling 
on the detector) at an off-axis angle of~$\approx 2^{\prime}$. 
Faint and very faint modes were used for the event telemetry format for our observations 
and the archival observations, respectively; 
\ASCA grade 0, 2, 3, 4 and 6 events were used in the analysis. 
Charge transfer inefficiency (CTI) problems do not significantly affect
the present data since all but one of the 
quasars were observed with the back-illuminated CCD S3. 
For the quasar SDSS~0210$-$0018, detected by the I2 front-illuminated CCD, 
the CTI was mitigated using the latest version of the corrector of Townsley et al. (2000). 

As a safety check, we have searched for background flares occurring 
during the observations. These features are due to ``space weather'' 
(primarily soft electrons interacting with the CCDs). 
The background level is constant to within 10--15\% in all the observations. 
The photon arrival times of all of the sources were also inspected 
to be certain that they were not produced or affected 
by ``cosmic ray afterglows'' (\CHANDRA X-ray Center 2000, private communication); 
note that cosmic ray afterglows are not typically seen on CCD S3. 
We are confident that the source parameters derived from the analysis 
described below are unaffected by transient spurious phenomena.

Source detection was carried out with {\sc wavdetect} (Dobrzycki et~al. 1999; 
Freeman et~al. 2001). 
For each image, we calculated wavelet transforms (using a Mexican hat kernel) 
with wavelet scale sizes of 1, 1.4, 2, 2.8, 4, 5.7, 8, 11.3 and 16 pixels. 
Those peaks whose probability of being false were less than the threshold 
of 10$^{-6}$ were declared real; 
detections were typically achieved for the smaller wavelet scales of 1.4 pixels or less 
as expected for these distant sources. 
Source searching was performed in four different energy ranges: the ultrasoft 
band (0.3--0.5 keV), the soft band \hbox{(0.5--2 keV)}, the hard band (2--8 keV) and the full band (0.5--8 keV); 
these correspond to the $\approx$ \hbox{1.7--2.8}, 2.8--11, 11--45 and 2.8--45 keV rest-frame bands 
(at the average redshift of \hbox{$z\approx$ 4.6} for the present sample). 
Nine of the 14 quasars are detected, and the {\sc wavdetect} photometry results are shown in Table~2. 
We have cross checked these results with manual aperture photometry; 
we found good agreement between the two techniques. 
Five sources went undetected even after increasing the false-positive threshold to 10$^{-4}$: 
SDSS~1129$-$0142, SDSS~1208$+$0010, PSS~1435$+$3057, SDSS~1532$-$0039, 
and SDSS~1605$-$0112. 
The SDSS~1208$+$0010 ($z=5.27$) full-band image shows two X-ray photons within 0.8\arcsec\/ of the optical position of the source. 
To investigate whether this source can be considered detected, we extracted a 
400$\times$400 pixel$^{2}$ region centered on the optical position of the quasar 
(we excluded the immediate vicinity of the quasar itself).  
This region was covered with 10000 circles of 1\arcsec\/ radius whose centers 
were randomly chosen.\footnote{The on-orbit 80\% encircled broad-band energy (EE) radius for on-axis sources is 0.685\arcsec\/ 
(see Fig.~6.3 of the \CHANDRA Proposers' Observatory Guide). This value has been summed in quadrature with 
the average positional errors of our clearly detected sources, i.e., the mean difference between the optical and 
X-ray positions ($\sim$ 0.6\arcsec). The derived radius is 0.9\arcsec\/, which we have approximated 
to 1\arcsec\/ in order to be reasonably conservative in the EE and positional error estimation. 
The PSF is quite stable in the sampled region due to its proximity to the optical axis, and the vignetting 
is negligible.}
The counts obtained in each circle were averaged, obtaining 0.0211 counts circle$^{-1}$. 
This value was subsequently checked by manually computing the value of the background around the 
quasar in the same extraction region, finding a very similar value (0.0226 counts circle$^{-1}$). 
The Poisson probability of obtaining two counts or more when 0.0211 counts 
are expected is \hbox{$\approx2.2$ $\times$ 10$^{-4}$}. 

\begin{figure*}[t]
{\vglue-2.0cm\centerline{\includegraphics[angle=0,width=23cm]{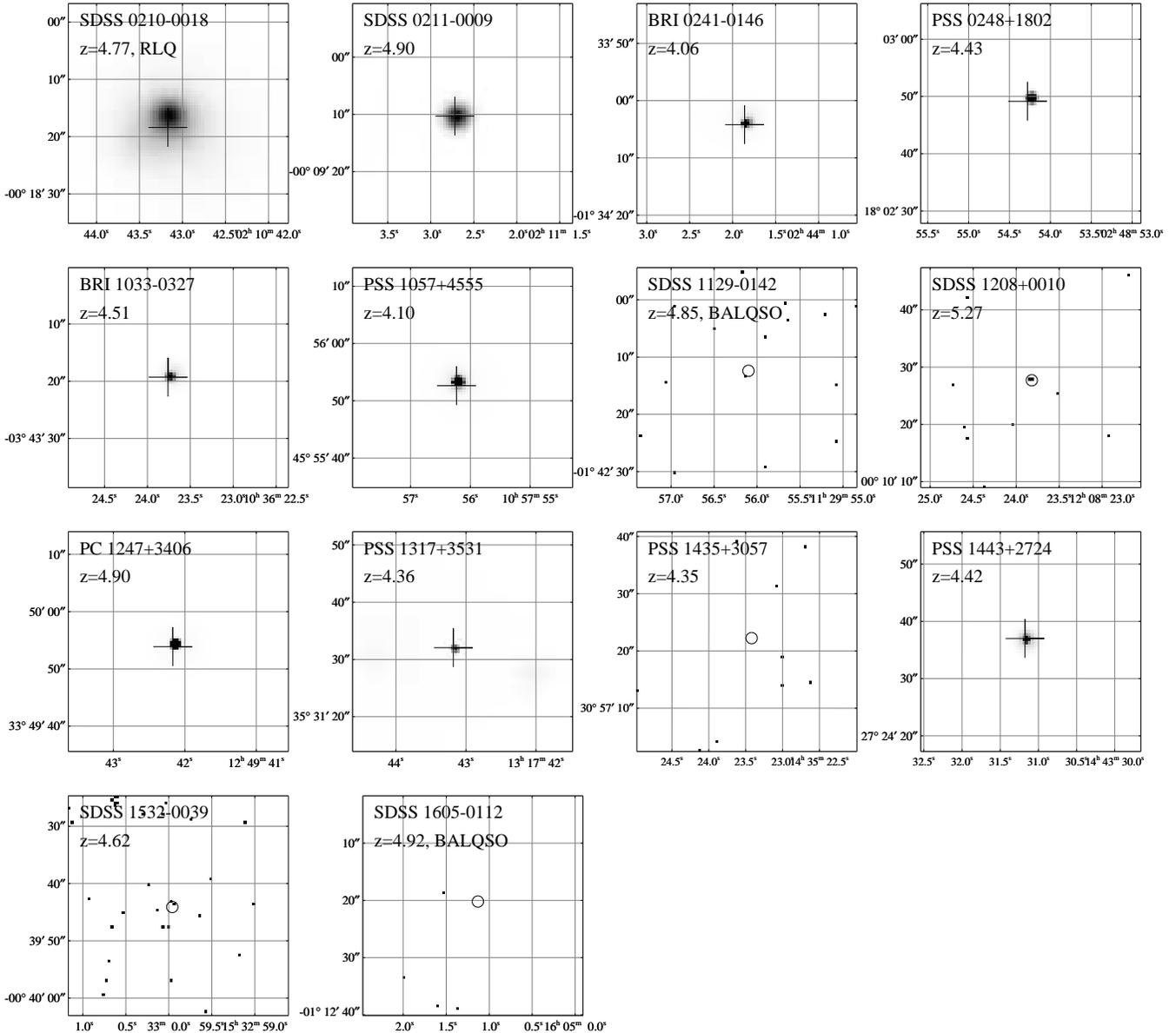}}}
\vglue-11.3cm\caption{\footnotesize \CHANDRA full-band images of the 14 high-redshift quasars. 
In each panel, the horizontal axis shows the Right Ascension, and the vertical axis shows the Declination 
(both in J2000 coordinates). 
Each box is 40\arcsec $\times$ 40\arcsec. 
For the nine sources with the best statistics, a 2$\sigma$ level of smoothing has been 
applied; for the other five no smoothing has been done. 
The optical position is indicated either by a cross or by a circle of 
1\arcsec \/ \/radius. For SDSS~0210$-$0018 there is a difference between the 
source peak and the X-ray centroid of the detection. This is probably due to the large off-axis angle 
($\sim$ 10.\arcmin3) at which the source has been detected.\label{fig1}}
\vglue-0.2cm
\end{figure*}

Recently, using extensive Monte-Carlo simulations on the \CHANDRA observation of the 
Hubble Deep Field North, Brandt et al. (2001b) 
found that the distribution function of background counts is almost Poisson. 
We have verified the validity of the Poisson approximation 
for the SDSS~1208$+$0010 field 
using the results of the 10000-circle analysis above. 
We obtained two counts or more in only two cases. 
Therefore, the derived probability of obtaining two counts or more is $\approx$ 2 
$\times$ 10$^{-4}$, which is similar to the value reported above. 
In the following, SDSS~1208$+$0010 will be considered detected by \CHANDRA 
(at about the 3.9\/$\sigma$ confidence level for a one-tailed Gaussian distribution). 

The situation is more uncertain for SDSS~1532$-$0039 \hbox{($z=4.62$)}, an interesting object which shows 
no emission lines in its optical spectrum (Fan et al. 1999b, 2000a). 
{\sc wavdetect} also reports no detection 
even though aperture photometry reveals two photons within 1.3\arcsec\/ of the 
optical position. The local background in this observation is 
higher than that for the SDSS~1208$+$0010 observation. 
By repeating the same procedure applied to the SDSS~1208$+$0010 field with 
an extraction circle of radius 1.3\arcsec\/, 
the average number of counts in each extraction circle is 0.0757, 
and the derived Poisson probability of 
obtaining two or more counts when 0.0757 are expected is \hbox{$\approx$ 2.7 $\times$ 10$^{-3}$.} 
The Monte-Carlo simulation is in agreement with this value ($\approx$ 3.2 $\times$ 10$^{-3}$). 

It is worth noting that for SDSS~1208$+$0010 and SDSS~1532$-$0039 
the two photons located near the optical position 
have not been obtained from a blind search across the entire \hbox{field-of-view}, 
but rather with ``a priori'' knowledge of the location of the sources. 
In the following only SDSS~1208$+$0010 will be treated as having been detected, due to its 
\hbox{order-of-magnitude} higher statistical significance. 

Three other sources have not been detected: 
the BALQSOs SDSS~1129$-$0142 ($z=4.85$) and SDSS~1605$-$0112 \hbox{($z=4.92$)}, 
and the quasar PSS~1435$+$3057 ($z=4.35$). 
Only one photon is present within 1.3\arcsec\/ of the optical position of SDSS~1129$-$0142, 
while no photons were detected for PSS~1435$+$3057 and SDSS~1605$-$0112 within 2\arcsec\/. 

\CHANDRA images from 0.5 to 8 keV of the quasars in the sample are shown in Fig.~1. 
The adaptive smoothing method of Ebeling, White, \& Rangarajan (2001) has been applied 
to the nine images of the sources with clear detections (using a 2$\sigma$ level of smoothing). 
\ For the other five quasars, \ the raw {\it Chan-}
\end{multicols}
\begin{deluxetable}{lccccccccccc}
\rotate
\tablecolumns{13}
\tabletypesize{\footnotesize}
\tablewidth{0pt}
\tablecaption{Properties of $z>4$ Quasars Observed by {\it CHANDRA}}
\tablehead{ 
\colhead{Object} & \colhead{$N_{\rm H}$\tablenotemark{a}} & 
\colhead{$AB_{1450(1+z)}$} & \colhead{$f_{2500}$\tablenotemark{b}} & \colhead{$\log (\nu L_\nu )_{2500}$} & 
\colhead{$M_B$} & \colhead{Count~rate\tablenotemark{c}} & 
\colhead{$f_{\rm x}$\tablenotemark{d}} & \colhead{$f_{\rm 2\/keV}$\tablenotemark{e}} & 
\colhead{$\log (\nu L_\nu )_{\rm 2\/keV}$} & \colhead{$\alpha_{\rm ox}$\tablenotemark{f}} & \colhead{$R$} \\
\colhead{(1)} & \colhead{(2)} & \colhead{(3)} & \colhead{(4)} & \colhead{(5)} & \colhead{(6)} & \colhead{(7)} &  
\colhead{(8)} & \colhead{(9)} & \colhead{(10)} & \colhead{(11)} & \colhead{(12)} 
}
\startdata
SDSS~0210$-$0018 & 2.66 & 19.3 & 1.06 & 46.3 & $-$26.8 & 5.65 & 28.1$^{+8.1}_{-6.0}$ & 
24.2 & 45.3 & $-$1.40$^{+0.07}_{-0.04}$ & 86.1--101.5\tablenotemark{g} \\
SDSS~0211$-$0009 & 2.60 & 20.0 & 0.56 & 46.1 & $-$26.2 & 1.01 & {\phn}3.1$^{+2.2}_{-1.4}$ & 
2.71 & 44.4 & $-$1.66$^{+0.12}_{-0.11}$ & 0.6\tablenotemark{h} \\
BRI~0241$-$0146 & 3.52 & 18.4 & 2.44 & 46.6 & $-$27.5 & 1.22 & {\phn}3.8$^{+1.9}_{-1.2}$ & 
2.89 & 44.3 & $-$1.89$^{+0.08}_{-0.09}$ & $<$1.6\tablenotemark{i} \\
PSS~0248$+$1802 & 9.18 & 18.1 & 3.21 & 46.8 & $-$27.9 & 5.20 & 19.6$^{+9.0}_{-6.3}$ & 
15.9 & 45.1 & $-$1.65$^{+0.08}_{-0.09}$ & 3.0\tablenotemark{i} \\
BRI~1033$-$0327 & 4.79 & 18.8 & 1.69 & 46.5 & $-$27.2 & 2.32 & {\phn}7.6$^{+3.9}_{-2.7}$ & 
6.25 & 44.7 & $-$1.70$\pm{0.09}$ & $<$0.9\tablenotemark{j} \\
PSS~1057$+$4555 & 1.11 & 17.6 & 5.09 & 46.9 & $-$28.3 & 7.48 & 21.6$^{+6.5}_{-4.9}$ & 
16.5 & 45.0 & $-$1.72$^{+0.06}_{-0.07}$ & 2.5 \\
SDSS~1129$-$0142 & 3.60 & 19.2 & 1.17 & 46.4 & $-$26.9 & $<$0.61 & $<$1.6 & 
$<$1.44 & $<$44.1 & $<$$-$1.88 & $<$3.3 \\
SDSS~1208$+$0010 & 2.00 & 20.5 & 0.35 & 45.9 & $-$25.8 & 0.43 & {\phn}1.1$^{+1.5}_{-0.7}$\tablenotemark{k} & 
1.02 & 44.0 & $-$1.74$^{+0.15}_{-0.19}$ & $<$13.1 \\
PC~1247$+$3406 & 1.30 & 19.2 & 1.17 & 46.4 & $-$26.9 & 2.99 & {\phn}8.7$^{+3.2}_{-2.4}$ & 
7.66 & 44.8 & $-$1.61$^{+0.09}_{-0.07}$ & $<$3.4 \\
PSS~1317$+$3531 & 0.99 & 18.9 & 1.54 & 46.4 & $-$27.1 & 0.76 & {\phn}2.2$^{+2.2}_{-1.2}$ & 
1.74 & 44.1 & $-$1.90$^{+0.13}_{-0.14}$ & $<$2.1\tablenotemark{i} \\
PSS~1435$+$3057 & 1.16 & 19.1 & 1.28 & 46.4 & $-$26.9 & $<$1.07 & $<$2.7 & 
$<$2.12 & $<$44.2 & $<$$-$1.83 & $<$2.1\tablenotemark{i} \\
PSS~1443$+$2724 & 2.12 & 19.0 & 1.40 & 46.4 & $-$27.0 & 3.22 & {\phn}9.7$^{+5.5}_{-5.1}$ & 
7.81 & 44.8 & $-$1.63$^{+0.09}_{-0.14}$ & $<$2.1\tablenotemark{i} \\
SDSS~1532$-$0039 & 4.60 & 19.4 & 0.97 & 46.3 & $-$26.7 & $<$1.24 & $<$3.4 & 
$<$2.86 & $<$44.4 & $<$$-$1.74 & $<$0.6\tablenotemark{h} \\
SDSS~1605$-$0112 & 9.07 & 19.4 & 0.97 & 46.3 & $-$26.8 & $<$0.65 & $<$2.0 & $<$1.77 & 
$<$44.2 & $<$$-$1.82 & $<$4.5 \\
\tableline
\enddata
\tablenotetext{a}{From Dickey \& Lockman (1990) in units of $10^{20}$ cm$^{-2}$ .}
\tablenotetext{b}{At rest-frame 2500~\AA\ in units of $10^{-27}$ erg cm$^{-2}$ s$^{-1}$ Hz$^{-1}$.}
\tablenotetext{c}{Observed count rate computed in the soft band, in units of $10^{-3}$  counts s$^{-1}$.}
\tablenotetext{d}{Galactic absorption-corrected flux in the observed 0.5--2 keV band in units 
of $10^{-15}$ erg cm$^{-2}$ s$^{-1}$.} 
\tablenotetext{e}{Rest-frame 2~keV flux density in units of $10^{-32}$ erg cm$^{-2}$ s$^{-1}$ Hz$^{-1}$.}
\tablenotetext{f}{Errors have been computed following the ``numerical method'' described in $\S$~1.7.3 of Lyons (1991); 
both the statistical uncertainties on the X-ray count rates 
and the effects of the observed ranges of the X-ray and optical continuum shapes have been taken into account.} 
\tablenotetext{g}{FIRST and NVSS report two different values for the flux density: 9.75$\pm{0.14}$ and 11.5$\pm{1.0}$ 
mJy, respectively.}
\tablenotetext{h}{1.4~GHz flux density from Carilli et al. (2001).}
\tablenotetext{i}{5~GHz flux density from Stern et al. (2000).}
\tablenotetext{j}{1.4~GHz flux density from Yun et al. (2000).}
\tablenotetext{k}{This flux has been computed from the full-band counts and then rescaled to the 0.5--2 keV band.}
\label{tab3}
\end{deluxetable}
\begin{multicols}{2}
\noindent {\it dra} images are shown. 
The optical position is indicated either by a cross or by a circle (for the faintest or undetected sources). 

To verify that source confusion problems are unlikely, 
we have computed the number of spurious associations expected at the conservative 0.5--2 keV 
flux limit of 10$^{-15}$ erg cm$^{-2}$ s$^{-1}$ assuming the integral source counts of Hasinger et al. (1998) 
and adopting an X-ray error circle radius of 1\arcsec \/ (see Table~1). 
We expect $\approx$ 2.4 $\times$ 10$^{-3}$ spurious associations summed over all of the 10 \CHANDRA detections. 

For the four sources with the most detected counts, 
SDSS~0210$-$0018, PSS~0248$+$1802, PSS~1057$+$4555 and PC~1247$+$3406, 
we computed a band ratio (BR), defined as the ratio between the \hbox{2--8 keV} and \hbox{0.5--2 keV} counts. 
We calculated errors at the $\approx1\sigma$ level for this quantity 
following the ``numerical method'' described in $\S$~1.7.3 of Lyons (1991); 
this avoids the failure of the standard approximate variance formula when the number of 
counts is small (see $\S$~2.4.5 of Eadie et al. 1971). 
For SDSS~0210$-$0018, PSS~1057$+$4555 and PC~1247$+$3406 we found BR=0.27$^{+0.15}_{-0.11}$, 0.29$^{+0.19}_{-0.13}$ 
and 0.21$^{+0.21}_{-0.05}$, respectively, which are consistent with unabsorbed $\Gamma\approx2$ power-law spectra. 
A flatter $\Gamma\approx1.5$ power-law slope, as is seen in many RLQs (e.g., Cappi et al. 1997), 
cannot be ruled out for SDSS~0210$-$0018 
given the statistical uncertainties on the source counts reported in Table~2. 

PSS~0248$+$1802, in contrast, is characterized by a larger band ratio (BR=0.67$^{+0.51}_{-0.34}$). 
This value corresponds to a power law with $\Gamma\approx1.2$. 
If a canonical $\Gamma=2$ is assumed, 
an intrinsic H~I column density larger than \hbox{5$\times$10$^{23}$ cm$^{-2}$} is derived. 
Given this possible absorption, 
the prominent soft ($E\simlt3$ keV in the rest frame) X-ray emission of PSS~0248$+$1802 appears surprising, 
unless a more complex model is adopted. For example, a partial covering model could 
provide a more reasonable parameterization of its X-ray counts; in this model 
a fraction of the nuclear radiation is able to escape unabsorbed. 
Although this model is clearly only a tentative and rough description of the 
PSS~0248$+$1802 X-ray spectrum, 
it is interesting to note that indications of complex \hbox{X-ray} absorbers have recently been 
obtained for several BALQSOs observed by \CHANDRA (Gallagher et al. 2001a) and for the BALQSO 
SDSS~1044$-$0125 detected by \XMM (B01; Mathur 2001). 
However, inspection of the optical spectrum (Fig.~3 in Kennefick et al. 1995) shows no evidence for UV BALs in PSS~0248$+$1802. 

We have not carried out a stacking analysis of the \CHANDRA RQQs; 
due to the small number 
of quasars thus far available, this analysis would be strongly biased toward 
the three RQQs (PSS~0248$+$1802, PSS~1057$+$4555 and PC~1247$+$3406) with the most detected counts.

\subsection{Simultaneous $I$-band photometry and spectroscopy}

Given that quasars frequently vary, sometimes spectacularly, in both the 
optical and X-ray bands, one must always be concerned about the reliability 
of the frequently employed \hbox{optical-to-X-ray} flux ratio whenever 
the observations in the two bands of a given quasar are not simultaneous. 
It is firmly established that there is an anticorrelation between 
variability and luminosity in the optical band 
(e.g., Cid Fernandes, Aretxaga, \& Terlevich 
1996 and references therein). In a recent study Kaspi (2001) shows 
that high-redshift, high-luminosity quasars ($2<z<3.4$) have longer 
optical variability time scales than low-redshift, low-luminosity 
quasars ($z<0.4$), and their variations over $\sim 1.5$ years (in the 
rest frame) amount to $\sim 10$\%. 
For the objects in this study, the observed-frame time between the 
optical and the \CHANDRA observations is about three years; we are 
helped by the fact that translating this observed temporal difference to the 
rest frame of the object reduces the delay by a factor of~$\approx$~5.6, 
so the typical time elapsed in the rest frame is several months. 

Although this is not a large time delay, we have been able to reduce the 
uncertainty in the optical luminosity at the time of the \CHANDRA observations 
by obtaining $I$-band images and 
\hbox{short-exposure}, low-resolution spectra within at most two weeks 
(a few days in the rest frame) for six of the seven quasars observed during Cycle~2. 
It has been possible to acquire 
timely optical data on these faint objects by utilizing the queue-scheduled 
nature of the Hobby-Eberly Telescope (HET; Ramsey et al. 1998). 

The data were taken with the HET's Marcario 
Low Resolution Spectrograph (LRS; Hill et al. 1998a,b; 
Cobos Duenas et al. 1998). 
When we were notified that 
a \CHANDRA observation of one of our quasars was to occur in the near future, 
we activated the object in the HET queue. 
At the next opportunity 
(seeing under 3$''$, reasonably clear but not necessarily photometric) a 
short (1--2 minute) $I$ image of the quasar field was taken, followed 
by an~$\approx15$ minute spectrum of the quasar. 

The image covered a field $\approx4\arcmin$ on a side, and using 
the published finding charts of the fields it was possible to obtain 
an approximate photometric calibration using stellar objects in the field. 
Although it is difficult to give precise estimates given the difference 
between the LRS and SDSS bandpasses, it is clear that none of the six quasars 
changed significantly (by $\simgt 30$\%) in brightness from their published values, and 
the spectra also show no significant changes from those published.

\section{Results}

In Table~3 the X-ray, optical and radio properties of the $z>4$ quasars observed by \CHANDRA 
are presented. 
A description is as follows. \\
{\sl Column (1)}. --- The abbreviated name of the source. \\
{\sl Column (2)}. --- The Galactic column density from Dickey \& Lockman (1990) in units of \hbox{10$^{20}$ cm$^{-2}$}. \\
{\sl Column (3)}. --- The monochromatic rest-frame $AB_{1450(1+z)}$ magnitude 
(with estimated errors of the order of 0.1 mag). \\
{\sl Columns (4) and (5)}. --- The 2500~\AA\ rest-frame flux density and luminosity. 
These were computed from the $AB_{1450(1+z)}$ magnitude assuming an optical power-law slope with 
$\alpha$=$-$0.79 \hbox{($S_{\nu}$ $\propto$ $\nu^{\alpha}$)}. 
This $\alpha$ value represents the estimate of the average spectral power-law slope in the rest-frame UV for quasars 
at \hbox{$z\approx4$} obtained by Fan et al. (2001), taking into account 
the selection completeness of their multicolor sample. Its corresponding 1$\sigma$ dispersion is 0.34. 
These values compare to $\alpha$=$-$0.91 with dispersion 0.26 from Schneider et al. (1991) and 
$\alpha$=$-$0.93 with dispersion 0.31 from Schneider et al. (2001). 
These measurements are all significantly steeper than the canonical $\alpha$=$-$0.5 median ($-$0.6 mean) value 
found by Richstone \& Schmidt (1980). Since the majority of their quasars have redshifts below 2, their measurements 
tend to be made at rest wavelengths considerably longer than those in the region 
where Fan et al. (2001) and Schneider et al. (2001) determined the continuum properties. 
Recently, Vanden Berk et al. (2001) derived a composite quasar spectrum from $\approx 2000$ SDSS quasars. 
Their best-fit spectral slope in the range $\approx$ 1300--5000 \AA\ (rest frame) is $-$0.44 
(with an uncertainty of $\approx 0.1$ mainly due to the 
spectrophotometric calibration of the spectra). 
The steeper indices 
\figurenum{2}
\centerline{\includegraphics[angle=0,width=8.5cm]{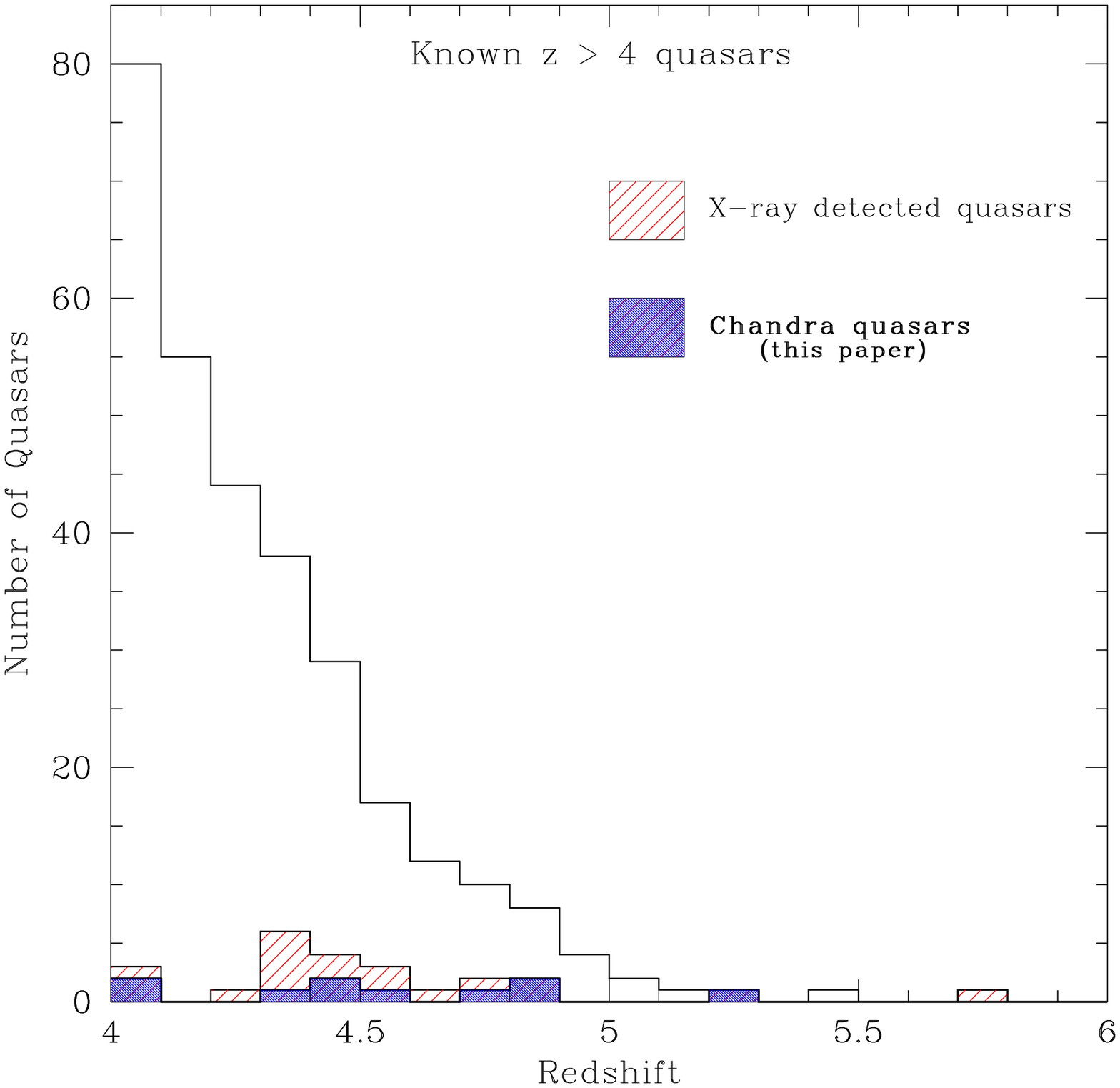}}
\figcaption{\footnotesize
Redshift distribution of all known $z>4$ quasars. 
The shaded regions indicate the 
quasars detected in the X-ray band, with particular emphasis on the $z>4$ \CHANDRA detections. 
\label{fig2}}
\centerline{}
\centerline{}
\noindent
obtained by Fan et al. (2001) and Schneider et al. (2001) may 
be due to the restricted wavelength range typically used in fitting the continua, 
as suggested by Schneider et al. (2001), and not to a change in the 
underlying spectral index at high redshift. 
We note that changing the optical continuum shape by \hbox{$\approx$ 0.3--0.4} 
has only a weak effect on $\alpha_{\rm ox}$ (see $\S$~4.2). \\
{\sl Column (6)}. --- The absolute $B$ magnitude calculated adopting $\alpha$=$-$0.79. 
Given the uncertainty in the continuum shapes of $z\approx4$ quasars between 
the rest-frame UV and optical bands (discussed above), a change from $\alpha$=$-$0.79 to 
$\alpha$=$-$0.5 induces an increase of $\approx40$\% in the inferred blue luminosity. \\
{\sl Columns (7) and (8)}. --- The observed count rate and the absorption-corrected 0.5--2 keV (observed-frame) flux. 
The 0.5--2 keV flux was computed using the source counts in the same band 
and the \CHANDRA X-ray Center Portable, Interactive, Multi-Mission Simulator (PIMMS) software (AO3 Version) 
for a power-law slope of 
$\Gamma=2$ ($\Gamma$=$-$$\alpha$$+$1; $N(E)\propto~E^{-\Gamma}$, where $N(E)$ is the number of photons). 
Samples of nearby RQQs are well fitted in the X-ray band by power-law continua 
with $\Gamma$=1.7--2.3 (e.g., George et al. 2000; Mineo et al. 2000; Reeves \& Turner 2000). 
Changing the X-ray spectral slope in the range $\Gamma$=1.7--2.3 
provides only a few percent change in the measured X-ray flux. 
As a further test, the 0.5--2 keV flux was also computed by converting the full-band counts 
in the same manner described above. The results are consistent with 
those tabulated in Table~3 to within 20\%. 
Because SDSS~0210$-$0018 was detected at a large off-axis angle, its count rate was corrected 
for vignetting. 
For the four undetected sources, X-ray upper limits (at the 95\% confidence level) were computed 
following Kraft, Burrows, \& Nousek (1991). \\
{\sl Columns (9) and (10)}. --- The rest-frame 2~keV flux density and luminosity. These were 
computed assuming the same power-law slope as in the count-rate-to-flux conversion. \\
{\sl Column (11)}. ---  The optical-to-X-ray power-law slope, $\alpha_{\rm ox}$, defined as 
\begin{equation}
\alpha_{\rm ox}=\frac{\log[(f_\nu (2\,{\rm keV})/f_\nu (2500\,{\mbox\AA})]}
{\log[\nu (2\,{\rm keV})/\nu (2500\,{\mbox\AA})]} \ \  
\end{equation}
where $f_\nu$ is the flux density at a given wavelength or energy. \\
{\sl Column (12)}. --- The radio-loudness parameter, defined as \hbox{$R$ = $f_{\rm 5~GHz}/f_{\rm 4400~\mbox{\scriptsize\AA}}$} 
(rest frame). 
The 5~GHz flux density was computed, unless otherwise stated (see 
the notes at the bottom of the table), from the NVSS (Condon et al. 1998) or the 
FIRST (Becker, White, \& Helfand 1995) 1.4~GHz flux density assuming a radio power-law slope of $\alpha=-0.8$. 
The upper limits reported in the table are at the 3\/$\sigma$ level. 
Strong radio-loud objects are usually characterized by $R>100$, whereas radio-quiet ones have $R<10$ 
(e.g., Kellermann et al. 1989). 
In this regard, the $R$ parameter for SDSS~0210$-$0018 ($R\approx$ 86--102) places it among the 
RLQ population. It represents the highest-redshift RLQ detected in the X-ray band. 
The previous highest-redshift X-ray detected RLQ was GB~1428$+$4217 at $z=4.72$ 
(Fabian et al. 1997). 
Note that the $R$ parameter of SDSS~0210$-$0018 is much smaller than those of the four $z>4$ blazars 
mentioned in $\S$~1; therefore, the fact that SDSS~0210$-$0018 has a significantly lower X-ray flux than these 
four blazars is not surprising.

\section{Discussion}

The capabilities of \CHANDRA for efficiently studying the high-redshift X-ray Universe 
are apparent from Fig.~1. 
With an average exposure time of $\approx4.2$~ks, 10 out of 14 objects have been detected 
in the $\approx$ 3--45 keV rest-frame band, thereby increasing the number of $z>4$ \hbox{X-ray} detected 
quasars by 71\% (see Fig.~2).\footnote{The full list of known $z>4$ quasars 
is available at http://www.astro.caltech.edu/$\sim$george/z4.qsos.}
Four of the five highest-redshift \hbox{X-ray} detections are presented here. 
Figure~3 shows that our \CHANDRA targets are uniformly distributed in 
$AB_{1450(1+z)}$ magnitude throughout the high-redshift quasar population. 
In the following, the X-ray and optical properties of the \CHANDRA objects will be 
compared with those of other high-redshift quasars, primarily the KBS sample ($\S$~4.1). 
In $\S$~4.2 the SEDs of the \CHANDRA quasars are discussed and 
compared with samples of optically selected RQQs in order to investigate 
any possible dependence of quasars' SEDs on redshift or optical luminosity. 
A discussion of the four quasars which were not detected by \CHANDRA is provided in $\S$~4.3.

\subsection{X-ray flux comparisons with other high-redshift quasars}

Figure~4 plots the unabsorbed 0.5--2~keV observed-frame flux versus 
$AB_{1450(1+z)}$ magnitude for the objects in our \CHANDRA sample 
as well as for objects from the KBS \ROSAT study and other recent investigations. 
Because this figure is constructed from {\it directly observed} 
physical quantities, it is robust (we present a complementary 
discussion based on $\alpha_{\rm ox}$, a parameter somewhat less directly
tied to observation, in the next section). 
KBS only reported \hbox{0.1--2~keV} fluxes for their quasars (to obtain the highest 
possible signal-to-noise ratio); 
to obtain \hbox{0.5--2~keV} fluxes for these objects we have re-analyzed 
the archival \ROSAT data with the \hbox{{\sc MIDAS/EXSAS}} package 
(Zimmermann et al. 1998) so that the \hbox{0.5--2~keV} fluxes could 
be determined directly from the counts in the exact same band.\footnote{We have also recalculated 
the 0.1--2 keV fluxes and obtained good agreement with KBS.}
This was important to obtain the most reliable 0.5--2~keV fluxes 
possible; converting the KBS 0.1--2~keV fluxes to 0.5--2~keV fluxes 
assuming a spectral shape is subject to errors in the assumed 
spectral shape, errors in the Galactic absorption correction, and 
\figurenum{3}
\centerline{\includegraphics[angle=0,width=8.5cm]{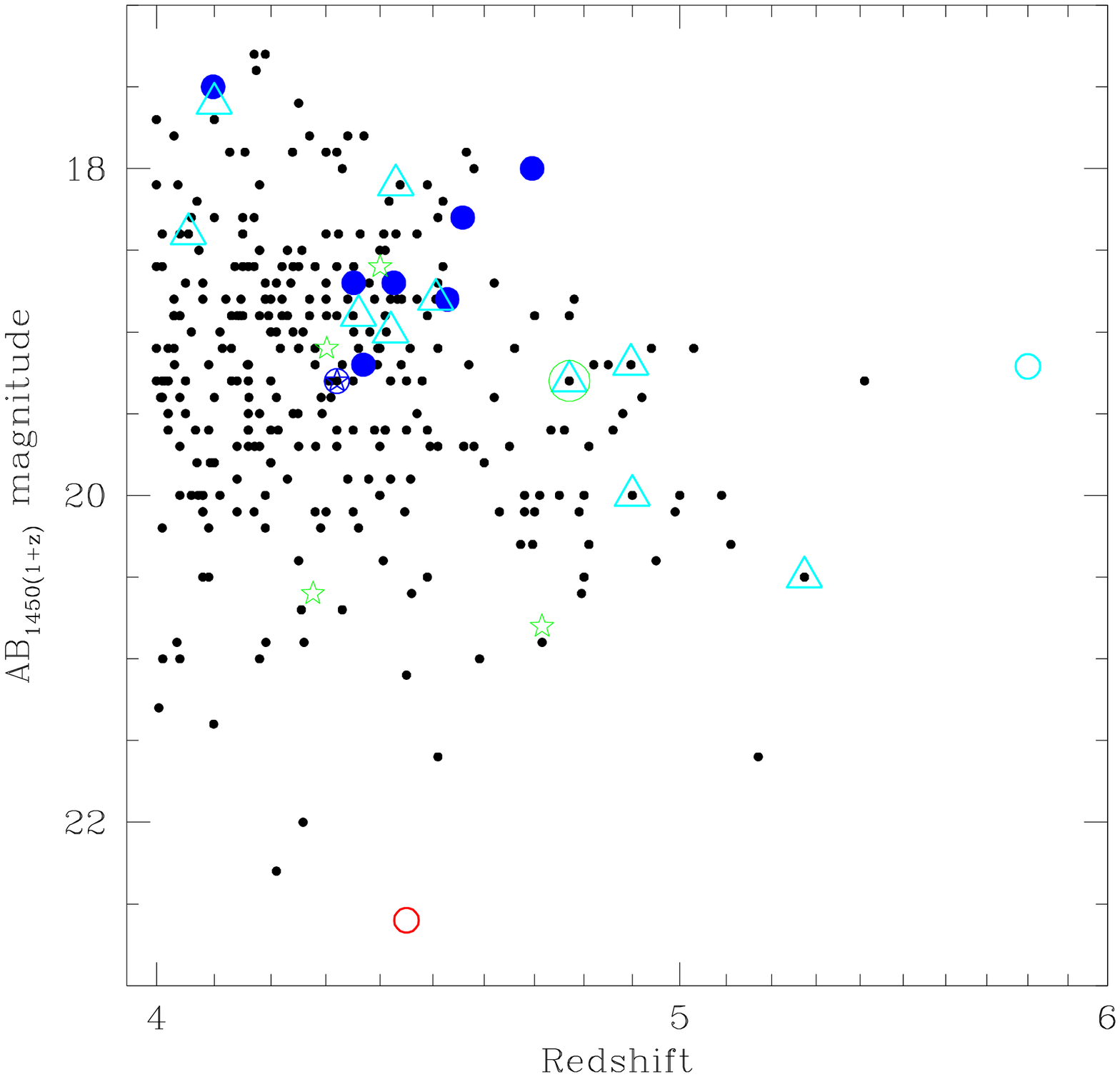}}
\figcaption{\footnotesize 
$AB_{1450(1+z)}$ magnitude versus redshift for all known $z>4$ quasars. 
For objects lacking a published value of $AB_{1450(1+z)}$, the magnitude has been 
calculated from the published magnitudes using empirical relationships between the 
observed fluxes in the different filters. 
RQQs with \CHANDRA detections are plotted as triangles, and 
the $z=4.77$ RLQ SDSS~0210$-$0018 is shown as a circled triangle. 
The {\it ROSAT}-detected RQQs and RLQ are plotted as filled circles and a circled star, respectively. 
Blazars from the KBS sample are plotted as open stars. 
For comparison, we have plotted as open circles the $z=4.45$ X-ray selected RQQ RX~J1052$+$5719 
(Schneider et al. 1998) and the $z=5.80$ RQQ SDSS~1044$-$0125 (Fan et al. 2000b; B01). 
The present \CHANDRA objects sample the $AB_{1450(1+z)}$ magnitude distribution well. 
Note the excellent capabilities of present X-ray instruments to detect 
high-redshift objects; in fact, the five highest-redshift objects so far detected in X-rays 
have been observed by \XMM (B01) and \CHANDRA (this paper). 
\label{fig3}}
\centerline{}
\centerline{}
\noindent
errors in the low-energy ($\simlt 0.5$~keV) calibration of 
the \ROSAT PSPC. Furthermore, in a few cases, additional archival 
observations have been used and averaged. We also used the 
{\sc XIMAGE} package (Giommi et al. 1992) to recalculate \ROSAT 
upper limits. All the results concerning the KBS objects presented 
in the following are derived from this re-analysis and are shown in Table~4. 

\begin{table*}[t]
\footnotesize
\caption{Properties of $z>4$ Quasars Originally Presented by KBS\label{tab4}}
\begin{center}
\begin{tabular}{lccccc}
\hline
\hline
Object & $z$ &  $AB_{1450(1+z)}$  & $M_B$ & $f_{\rm x}$\tablenotemark{a} & 
$\alpha_{\rm ox}$ \\
\hline
\cutinhead{\ROSAT detections}
Q~0000$-$2619 & 4.10 & 17.5 & $-$28.4 & 2.55 & $-$1.71 \\
BR~0019$-$1522 & 4.53 & 18.8 & $-$27.2 & 1.89 & $-$1.56 \\
BR~0351$-$1034 & 4.35 & 18.7 & $-$27.3 & 1.62 & $-$1.60 \\
BR~0951$-$0450 & 4.37 & 19.2 & $-$26.8 & 1.44 & $-$1.54 \\
BRI~0952$-$0115 & 4.43 & 18.7 & $-$27.3 & 2.45 & $-$1.52 \\
BR~1202$-$0725 & 4.70 & 18.0 & $-$28.1 & 2.29 & $-$1.63 \\
RX~J1759.4$+$6638 & 4.32 & 19.3 & $-$26.7 & 1.35 & $-$1.53 \\
BR~2237$-$0607 & 4.56 & 18.3 & $-$27.7 & 2.49 & $-$1.58 \\
\cutinhead{\ROSAT Upper Limits}
PC~0027$+$0525 & 4.10 & 21.4 & $-$24.5 & $<$8.93 & $<$$-$0.90 \\
PC~0027$+$0521 & 4.21 & 22.3 & $-$23.6 & $<$8.69 & $<$$-$0.77 \\
Q~0046$-$293 & 4.01 & 19.3 & $-$26.5 & $<$5.65 & $<$$-$1.31 \\
Q~0051$-$279 & 4.40 & 19.2 & $-$26.8 & $<$2.38 & $<$$-$1.45 \\
Q~0101$-$304 & 4.07 & 20.0 & $-$25.9 & $<$0.86 & $<$$-$1.51 \\
BRI~0103$+$0032 & 4.43 & 18.8 & $-$27.2 & $<$3.51 & $<$$-$1.45 \\
SDSS~033829.31$+$002156.3 & 5.00 & 20.0 & $-$26.2 & $<$8.55 & $<$$-$1.10 \\
PC~0953$+$4749 & 4.46 & 19.1 & $-$26.9 & $<$3.28 & $<$$-$1.41 \\
BRI~1050$-$0000 & 4.29 & 19.4 & $-$26.5 & $<$4.24 & $<$$-$1.33 \\
BR~1144$-$0723 & 4.15 & 18.8 & $-$27.1 & $<$2.39 & $<$$-$1.52 \\
SDSS~122600.68$+$005923.6 & 4.25 & 19.1 & $-$26.8 & $<$4.20 & $<$$-$1.38 \\
PC~1233$+$4752 & 4.45 & 20.1 & $-$25.9 & $<$8.02 & $<$$-$1.11 \\
PKS~1251$-$407 & 4.46 & 19.7 & $-$26.3 & $<$8.99 & $<$$-$1.15 \\
SDSS~131052.52$-$005533.4 & 4.14 & 18.9 & $-$27.0 & $<$2.40 & $<$$-$1.51 \\
Q~2133$-$4311 & 4.26 & 21.1 & $-$24.8 & $<$3.37 & $<$$-$1.11 \\
Q~2139$-$4324 & 4.46 & 20.6 & $-$25.4 & $<$4.30 & $<$$-$1.14 \\
PC~2331$+$0216 & 4.09 & 19.8 & $-$26.3 & $<$5.28 & $<$$-$1.24 \\
\cutinhead{Blazars}
RX~J1028.6$-$0844 & 4.28 & 20.6 & $-$25.3 & 36.2 & $-$0.79 \\
GB~1428$+$4217 & 4.72 & 20.8 & $-$25.3 & 48.9 & $-$0.69 \\
GB~1508$+$5714 & 4.30 & 19.1 & $-$26.8 & 56.4 & $-$0.94 \\
PMN~J0525$-$3343 & 4.40 & 18.6 & $-$27.4 & 21.8 & $-$1.18 \\
\hline
\end{tabular}
\vskip 2pt
\parbox{4.0in}
{\small\baselineskip 9pt
\footnotesize
\indent
$\rm ^a$ Galactic absorption-corrected flux in the observed 0.5--2 keV band in units 
of $10^{-14}$ erg cm$^{-2}$ s$^{-1}$. \\
}
\end{center}
\vglue-0.7cm
\end{table*}
\normalsize
Inspection of Fig.~4 shows that 
the \CHANDRA $z>4$ quasars tend to lie below the \ROSAT ones by a typical factor of $\approx3$ in X-ray flux 
(even within the same optical flux range). 
Only three \CHANDRA objects, the $z=4.77$ RLQ SDSS~0210$-$0018, the $z=4.10$ RQQ PSS~1057$+$4555 and the $z=4.43$ 
RQQ PSS~0248$+$1802, have X-ray fluxes comparable to those of the \ROSAT quasars. 
Comparing the soft X-ray fluxes of the 
\CHANDRA and \ROSAT objects with $AB_{1450(1+z)}$ magnitudes in the range $\approx$ 17.5--19.5
(excluding the BALQSOs SDSS~1129$-$0142 and SDSS~1605$-$0112 and the \ROSAT non-detections)
by means of the Kolmogorov-Smirnov test, we obtain 
that the probability that the two samples are drawn from 
the same parent population is $\approx1.8$ $\times$ 10$^{-4}$. 
However, it is possible that some of the undetected KBS quasars in 
the above range of $AB_{1450(1+z)}$ magnitude are as X-ray faint as the \CHANDRA objects. 

Several checks have been carried out in order to try to understand this difference. 
We have first looked for an instrumental origin. At the present time, 
the \CHANDRA ACIS calibration is uncertain to $\approx15$\% at energies 
below $\approx1$ keV (R. Kilgard 2001, private communication), 
mainly due to potential problems in 
\figurenum{4}
\centerline{\includegraphics[angle=0,width=8.5cm]{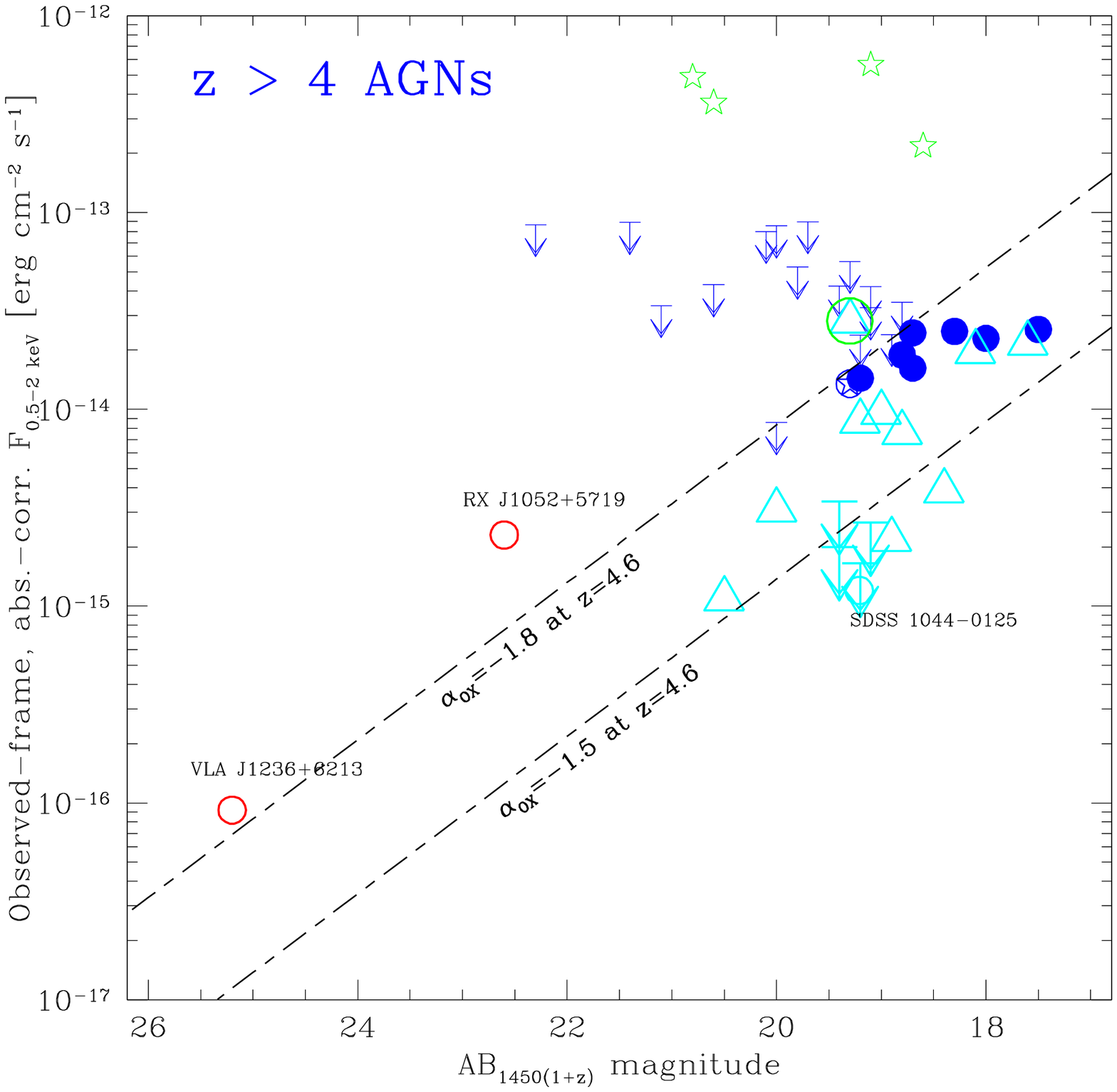}}
\figcaption{\footnotesize Observed-frame, Galactic absorption-corrected 0.5--2 keV flux versus $AB_{1450(1+z)}$ 
magnitude for $z>4$ AGNs. 
Symbols are as in Fig.~3. Small arrows indicate \ROSAT upper limits, while 
\CHANDRA upper limits are shown as large arrows. 
For comparison, we have plotted as open circles the $z=4.424$ Seyfert galaxy VLA J1236+6213 
(Waddington et al. 1999; Brandt et al. 2001b), the $z=4.45$ RQQ RX~J1052$+$5719, and the 
$z=5.80$ RQQ SDSS~1044$-$0125. 
The slanted lines show $\alpha_{\rm ox}=-1.5$ and $\alpha_{\rm ox}=-1.8$ loci for 
$z=4.6$ (the average redshift of the present sample), assuming the same X-ray 
and optical spectral shapes used in the text. 
\label{fig4}}
\centerline{}
\centerline{}
\noindent the ACIS quantum efficiency at low energies. 
A mis-calibration 
between the \ROSAT PSPC and \CHANDRA ACIS 
could be present as well, but it is very unlikely to be large enough to account for 
the observed difference of a factor of $\approx 3$ in the average flux. 
A physical explanation for these findings thus appears more likely, although we 
note that significant evolution would be surprising given the small difference 
in the average redshifts of the two samples (\hbox{$\Delta$$z\approx0.2$}, 
corresponding to a time interval of \hbox{$\approx40$ Myr}). Systematically
different selection effects for the KBS objects and the \CHANDRA objects may well
play a role in causing the observed X-ray flux difference. For example, the
SDSS could be more effective at finding objects that suffer from X-ray absorption
or have intrinsic SEDs with weaker X-ray emission. To look for
obvious selection effects, we have compared the optical emission-line properties 
of the \CHANDRA quasars with those of KBS; the optical emission lines could be
affected by reddening if absorption is present, or they could respond to a 
different ionizing continuum shape. No obvious correlation of the X-ray properties 
with the intensities of the Ly\/$\alpha$ or \ion{C}{4} emission lines is apparent, 
but the non-uniform spectral coverage of the quasars prevents us from carrying out
a quantitative analysis of this issue. In addition, there are not sufficient 
data to search effectively for a systematic difference in the optical continuum
shape. 

For the \CHANDRA sample, a correlation of the X-ray flux with the $AB_{1450(1+z)}$ magnitude 
is suggested by Fig.~4, where three out of the five X-ray faintest \CHANDRA quasars 
(excluding the BALQSOs, which will be discussed in $\S$~4.3) also have the faintest optical $AB_{1450(1+z)}$ magnitudes. 
Correlations of the \hbox{X-ray} flux with the optical magnitude for quasars have been extensively studied at low redshift 
(e.g., Zamorani et al. 1981). 
Applying the Spearman rank-order correlation test to the \CHANDRA objects excluding the RLQ and the BALQSOs, we find 
that the correlation of the X-ray flux with the $AB_{1450(1+z)}$ magnitude is significant at the $\approx97.3$\% level. 
Including also the RQQs of the KBS sample (with the upper limits), the correlation becomes significant 
at the $\approx99.9$\% level.

\subsection{Comparison of the Spectral Energy Distributions}

To investigate whether high-redshift quasars have different SEDs with respect to 
lower-redshift samples, Fig.~5 plots the $\alpha_{\rm ox}$ index against the luminosity density at 2500~\AA\ 
for radio-quiet, optically selected samples of quasars. Following Pickering, Impey, \& Foltz 
(1994) and B01, as comparison samples we used the 
Bright Quasar Survey (BQS; Schmidt \& Green 1983) RQQs and the 
Large Bright Quasar Survey (LBQS; Hewett, Foltz, \& Chaffee 1995) RQQs; the latter provides a well-defined 
comparison sample with a significantly higher median luminosity than that of the BQS. 
Further details can be found in B01 and in the caption of Fig.~5. 
We have recalculated both the 2500~\AA\ luminosity densities and the $\alpha_{\rm ox}$ values 
for the KBS sample using the same optical continuum adopted in this paper. 

Obviously, the large scatter of the data points in Fig.~5, coupled with the relatively small number 
of high-redshift quasars thus far detected in the X-ray band, prevents us from drawing firm conclusions about an 
evolutionary trend of the SED. 
However, the KBS quasars are characterized, on average, 
by SEDs which resemble those of lower-redshift samples reasonably well. 
In contrast, the \CHANDRA quasars appear to populate preferentially the region of lower $\alpha_{\rm ox}$ values. 
The postulated steepening of $\alpha_{\rm ox}$ with increasing optical luminosity 
(e.g., Pickering et al. 1994; Avni, Worrall, \& Morgan 1995 and references therein) 
cannot explain these results easily, since the SDSS objects have lower optical luminosities on average than the KBS sample. 

This finding can be checked in a more quantitative way by computing the average $\alpha_{\rm ox}$, 
taking into account the upper limits. 
For this purpose, we have used the ASURV software package Rev~1.2 (LaValley,
Isobe, \& Feigelson 1992). 
The optically selected RQQs observed by \ROSAT (KBS) have $\langle\alpha_{\rm ox}\rangle$ = $-$1.58$\pm{0.03}$ 
(the quoted errors represent the standard deviation of the mean), 
similar to the results obtained from the $z<0.5$ BQS excluding the Seyfert galaxies 
and the absorbed objects (\hbox{$\langle\alpha_{\rm ox}\rangle$ = $-$1.56$\pm{0.02}$}). 
The \CHANDRA $z>4$ quasars, by comparison, are characterized by a steep, more negative value of $\alpha_{\rm ox}$, 
with an average value of $-$1.78$\pm{0.03}$ ($-$1.75$\pm{0.03}$ if we exclude the two undetected BALQSOs). 
They lie between the \ROSAT and BQS samples on one hand and 
the value obtained by B01 ($\alpha_{\rm ox}=-1.91$) for the X-ray weak \hbox{$z=5.80$} quasar on the other. 
For this object intrinsic and/or associated 
\figurenum{5}
\centerline{\includegraphics[angle=0,width=8.5cm]{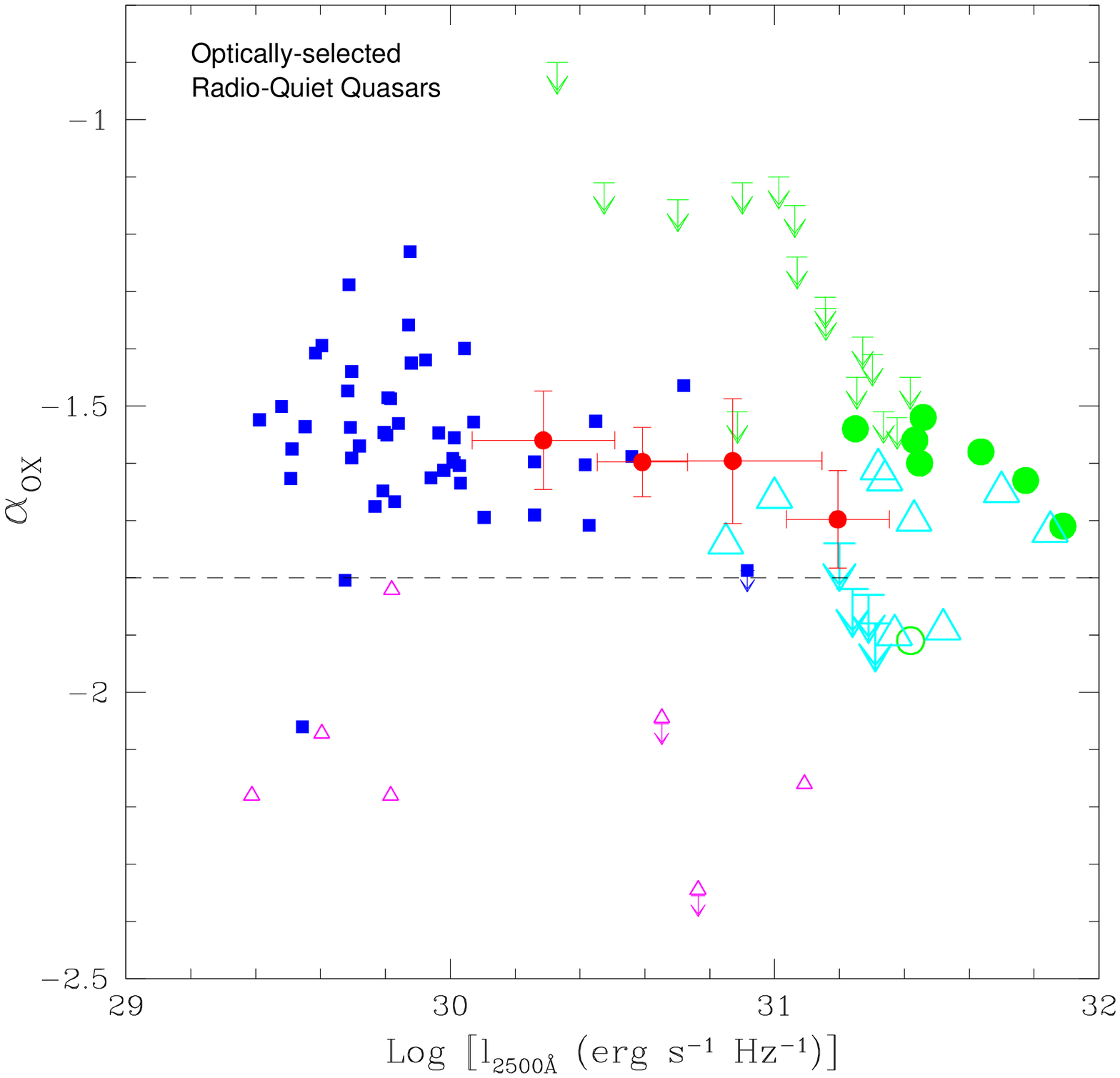}}
\figcaption{\footnotesize
$\alpha_{\rm ox}$ versus luminosity density at 2500~\AA\ for optically selected RQQs. In particular: 
(1) the present \CHANDRA quasars (large triangles and large upper limits); 
(2) the optically selected $z>4$ RQQs from KBS (large solid circles for the 
X-ray detected objects and small downward pointing arrows for the upper limits); 
(3) the $z=5.80$ SDSS X-ray weak quasar from B01 (open circle); 
(4) the seven luminous BQS RQQs (Schmidt \& Green 1983) 
whose $\alpha_{\rm ox}$ values and other properties indicate 
the presence of intrinsic absorption (small triangles, the upper limits being 
the quasars 0043$+$039 and 1700$+$518); 
(5) the other 46 luminous BQS RQQs (solid squares, one with 
a downward pointing arrow representing an upper limit for the quasar 1259$+$593); 
(6) the LBQS RQQs from Fig.~6b of Green et~al. (1995; filled circles with error bars). 
The LBQS RQQ data points were derived using an X-ray counts stacking 
technique as described in Green et~al. (1995); from left to right, these 
data points represent 21, 21, 35 and 70 LQBS RQQs, respectively. 
The three solid squares notably below the general trend 
(with $\alpha_{\rm ox}\la -1.8$, dashed line) are, from left to right, 
2214$+$139, 1552$+$085 and 1259$+$593 
(see B01 and Brandt, Laor, \& Wills 2000 for detailed discussion).
\label{fig5}}
\centerline{}
\centerline{}
\noindent X-ray absorption has been suggested, and recently near-infrared 
spectroscopy has revealed its BALQSO nature (Maiolino et al. 2001; R.~W. Goodrich et al., in preparation). 
Similarly, X-ray absorption may be present in some of the \CHANDRA quasars studied here. 

The assumption of a different optical power-law continuum (e.g., $\alpha$=$-$0.5; Richstone \& Schmidt 1980) 
results in a slightly flatter SED for the \CHANDRA sources \hbox{($\langle\alpha_{\rm ox}\rangle$ = $-$1.75$\pm{0.03}$)}. 
Furthermore, if we assume different X-ray spectral shapes of \hbox{$\Gamma=1.7$} and \hbox{$\Gamma=2.3$} 
(with the optical slope $\alpha$=$-$0.79 adopted in the paper), 
we find 
\hbox{$\langle\alpha_{\rm ox}\rangle$ = $-$1.83$\pm{0.03}$} and \hbox{$\langle\alpha_{\rm ox}\rangle$ = $-$1.73$\pm{0.03}$}, 
respectively. 

Figure~6 shows $\alpha_{\rm ox}$ versus redshift for $z>4$ \ROSAT (KBS), \XMM (B01) and 
the present \CHANDRA RQQs. 
For comparison, $\alpha_{\rm ox}$ versus redshift for lower-redshift, optically selected 
RQQs from the BQS and LBQS (Fig.~6d from Green et al. 1995) is also plotted. 
There is no clear evidence for a systematic 
decrease of $\alpha_{\rm ox}$ with redshift in the range $z\approx4$--6. This is confirmed by 
the Spearman test. 
However, the small number of \CHANDRA quasars at the highest redshifts (only three have been detected 
at \hbox{$z\ga4.8$}) prevents us from drawing firm conclusions about any evolutionary trend. 
This issue clearly requires further investigation with larger samples of radio-quiet objects at high redshift, 
which will be provided by the SDSS in the next few years for X-ray follow-up with \CHANDRA and {\it XMM-Newton}.

\subsection{Chandra undetected quasars}

The optical spectra of the 10 quasars detected by \CHANDRA show no evidence of unusual properties. 
In contrast, three out of the four undetected objects, SDSS~1129$-$0142, SDSS~1532$-$0039 and SDSS~1605$-$0112, 
are characterized by peculiar optical (UV rest frame) properties. 
SDSS~1129$-$0142 (Zheng et al. 2000) and SDSS~1605$-$0112 (Fan et al. 2000a) are BALQSOs. 
\hbox{X-ray} studies at low redshift show that absorption plays a major role in hiding BALQSOs 
in the \hbox{X-ray} band (e.g., Green \& Mathur 1996; Gallagher et al. 1999, 2001a,b; Brandt, Laor, \& Wills 2000; 
Green et al. 2001). 
Typical H~I column densities for high-ionization \hbox{BALQSOs} are 
\hbox{$\approx$ (1--10)$\times$10$^{22}$ cm$^{-2}$}, while for low-ionization BALQSOs 
the column densities may be \hbox{$\ga5\times10^{23}$ cm$^{-2}$}. 
Unfortunately, due to the limited optical spectral coverage at the present time, it is not possible 
to determine whether SDSS~1129$-$0142 and SDSS~1605$-$0112 are low-ionization BALQSOs. 
Even taking into account the sharp decrease of ACIS effective area above $\approx5$ keV, 
the energy band sampled by \CHANDRA corresponds to the \hbox{$\approx$ 2--30 keV} rest-frame energy range. 
Therefore the non-detections of SDSS~1129$-$0142 and SDSS~1605$-$0112, 
coupled with their $\alpha_{\rm ox}$ upper limits of $-$1.88 and $-$1.82 respectively, 
make them good candidates to be highly-absorbed objects, perhaps being Compton thick. 
Assuming the average $\alpha_{\rm ox}$ of our sample excluding the BALQSOs ($-$1.75) and a typical 
X-ray slope of $\Gamma=2.0$, we infer that a non-detection is achieved for a neutral column density of 
$\ga5$ $\times$ 10$^{23}$ cm$^{-2}$ for both sources. 
It is notable that SDSS~1044$-$0125 (B01), SDSS~1129$-$0142 and SDSS~1605$-$0112 all appear to have large 
X-ray absorbing column densities more characteristic of those seen in low-ionization BALQSOs 
at low redshift. Since it is unlikely a priori that all three of these objects are low-ionization BALQSOs 
(assuming, as at low redshift, that only $\sim1/10$ of \hbox{BALQSOs} are low-ionization BALQSOs), this may suggest 
that the X-ray absorption in high-ionization \hbox{BALQSOs} increases with redshift. 
Clearly further observations are required before firm conclusions can be drawn. 

Another quasar undetected in our observations is SDSS~1532$-$0039 (Fan et al. 1999b, 2000a), 
whose optical spectrum is emission-line free.
Its redshift of 4.62 has been obtained by identifying 
the two highly significant breaks at 6800 and 5100 \AA\ 
as the onset of the Ly$\alpha$ forest and a Lyman limit system, respectively. 
The $\alpha_{\rm ox}$ upper limit for this object of $-$1.74 marks it as 
moderately X-ray weak, although it need not be an extreme object. 
If X-ray absorption is present and is depressing the $\alpha_{\rm ox}$ value, corresponding absorption 
in the rest-frame UV could explain the lack of observed emission lines 
(such a phenomenon is observed in some BALQSOs; e.g., Wills, Brandt, \& Laor 1999). 
At present there is limited coverage of the \ion{C}{4} line for this object; 
obtaining a better signal-to-noise ratio 
spectrum would allow a detailed search for broad absorption lines from outflowing matter. 
We note that the relatively weak X-ray emission is not consistent with that 
expected from a BL Lac; this further supports the arguments of Fan et al. (1999b) that this object 
is unlikely to be a BL Lac. 
Anderson et al. (2001) have recently discovered two further $z>4$ quasars with anomalously weak emission lines; 
it will be interesting to determine if these objects are X-ray weak. 

The optical spectrum of the last undetected quasar, PSS~1435$+$3057, 
shows no obvious BAL or other absorption 
\begin{figure*}[t]
\figurenum{6}
\vglue-4.5cm
\centerline{\includegraphics[angle=0,width=18.5cm]{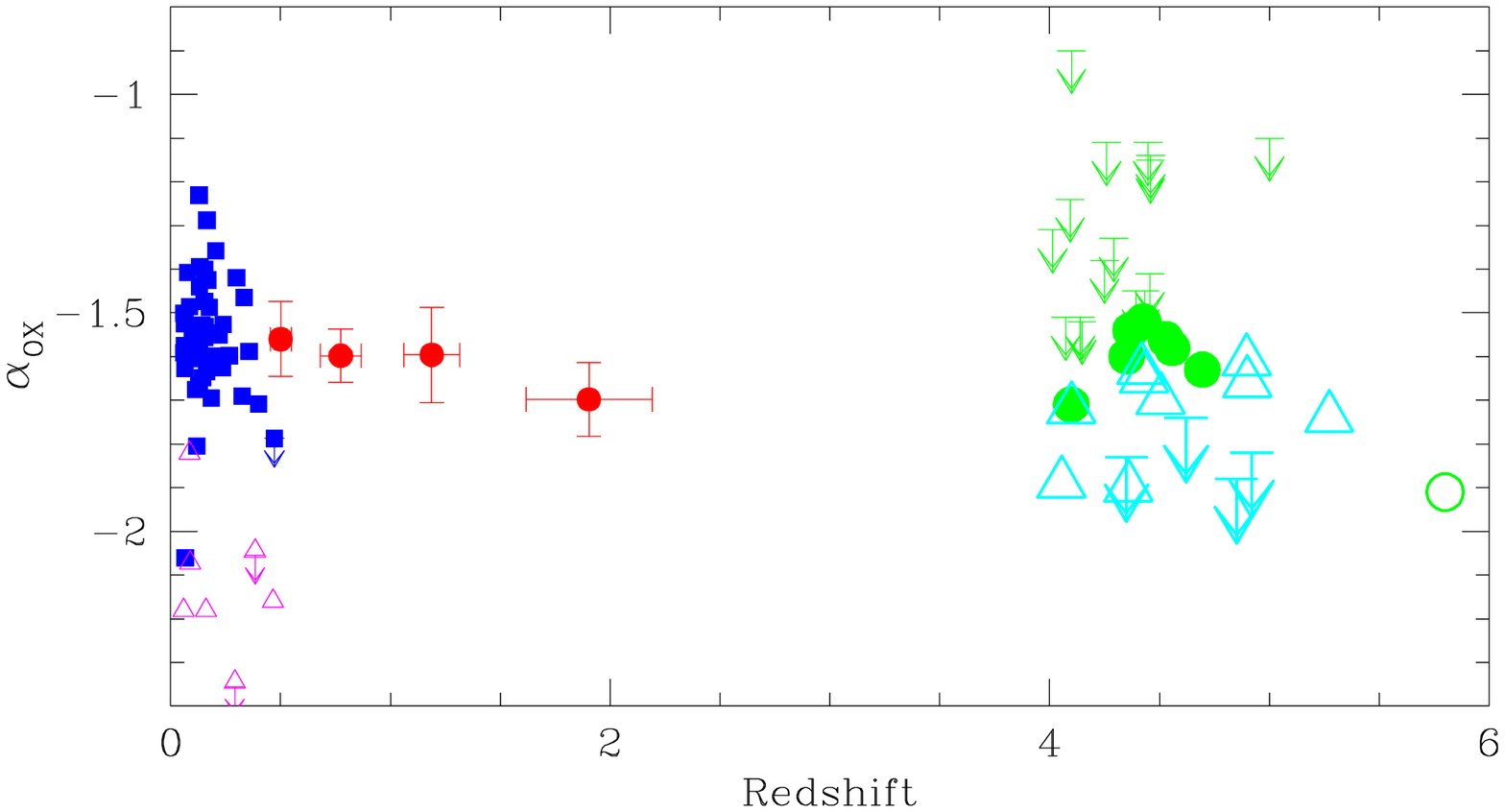}}
\vglue-3.8cm
\caption{\footnotesize $\alpha_{\rm ox}$ versus redshift for the optically selected RQQs 
observed by \ROSAT (KBS, large solid circles and thin upper limits), 
\CHANDRA (this paper, large triangles and thick upper limits), and for the 
$z=5.80$ RQQ observed by \XMM (B01, large open circle). 
For comparison, low-redshift RQQs are also plotted: 
the seven luminous and probably absorbed BQS RQQs (small open triangles), 
the other 46 luminous BQS RQQs (small solid squares), 
and the LBQS RQQs from Fig.~6d of Green et~al. 
(1995; small filled circles with error bars, see the caption of Fig.~5 for further details). 
There is no evidence for a systematic decrease of $\alpha_{\rm ox}$ with increasing redshift 
for $4<z<6$. 
\label{fig6}}
\end{figure*}
features, although the signal-to-noise ratio in the optical spectrum 
is low (Kennefick et al. 1995). 
It should be noted that this RQQ was observed for only 2.8~ks by \CHANDRA.

\section{Revealing the high-redshift Universe through X-rays}

We have found that some of the \CHANDRA $z>4$ quasars 
are moderately X-ray faint given their optical luminosities, by a factor of $\approx2$ 
with respect to the BQS sample. 
This effect may be due to the presence of 
large amounts of gas in the quasars' host galaxies fueling the active nuclei. 
In this regard, X-ray absorption has been detected in many $z\ga2$ RLQs and appears to increase with redshift 
(e.g., Elvis et al. 1994; Cappi et al. 1997; Fiore et al. 1998; Reeves \& Turner 2000). 
That absorption could play a role in obscuring high-redshift quasars was 
also suggested to explain the weak X-ray emission of the 
$z=5.80$ quasar SDSS~1044$-$0125 ($\alpha_{\rm ox}=-1.91$; B01), which indeed 
was revealed to be a BALQSO (Maiolino et al. 2001; R.~W. Goodrich et al., in preparation). 
Evidence for large amounts of gas and possibly dust in high-redshift quasars has been obtained for 
about one-third of the objects with multiwavelength data available. 
Indeed, in their study of optically selected $z>4$ quasars from the APM sample 
(Irwin, McMahon, \& Hazard 1991), Omont et al. (1996a) detected 
240~GHz continuum emission in six out of 16 quasars. Recently Omont et al. (2001) 
found 1.2~mm emission from 18 out of 62 $z\ga3.8$ quasars selected from the Palomar Sky Survey 
(Djorgovski et al. 1998).  
Follow-up observations 
of high-redshift quasars have revealed CO emission in several cases 
(Ohta et al. 1996, 1998; Omont et al. 1996b; Guilloteau et al. 1997, 1999; Carilli, Menten, \& Yun 1999). 
If the sub-mm/far-infrared emission is largely due to warm dust in the quasar host system, then 
molecular gas and dust masses of the order of 10$^{11}$ and \hbox{10$^{8}$ M$_{\odot}$}, respectively, are implied 
in most of these systems (Benford et al. 1999; Carilli et al. 1999). 
Absorption by this gas 
may be responsible for the X-ray faintness of some of the \CHANDRA quasars as well. 
However, the situation concerning the growth phase of 
massive black holes and their X-ray properties is probably complex and varied. 
To date there is no clear correlation between 1.2~mm detections of QSOs and their X-ray faintness; 
indeed, 1.2~mm observations failed to detect 
emission from four \CHANDRA objects (SDSS~0211$-$0009 and SDSS~1532$-$0039; Carilli et al. 2001; 
PSS~1435$+$3057 and PSS~1443$+$2724; Omont et al. 2001), 
while three quasars in our sample, BRI~1033$-$0327, PSS~1057$+$4555 and PSS~1317$+$3531, 
were successfully detected (Omont et al. 1996a, 2001; Guilloteau et al. 1999). 

The X-ray faintness of \CHANDRA high-redshift quasars could be alternatively explained 
by a highly super-Eddington accretion rate 
for which ``trapping radius'' effects cause the X-rays created in the inner regions of the accretion 
flow to be dragged back into the black hole (e.g., Begelman 1978; Rees 1978). 
The accretion flow would have a luminosity of about the Eddington limit even though the mass accretion rate 
is super-Eddington. 
The low radiative efficiency associated with an accretion flow 
with trapping effects would allow the rapid growth of a massive 
black hole in the early Universe. 

While the present observations directly show that \CHANDRA and \XMM can substantially increase 
the number of \hbox{$z\approx$ 4--5.3} quasars detected in the X-ray band, detailed spectroscopic 
studies of typical quasars at the highest redshifts 
\figurenum{7}
\centerline{\includegraphics[angle=-90,width=8.5cm]{vignali.fig7a.ps}}
\centerline{\includegraphics[angle=-90,width=8.5cm]{vignali.fig7b.ps}}
\figcaption{\footnotesize Simulated 40~ks \XEUS 
(final configuration, top panel) and {\it Constellation-X} (bottom panel) spectra of the $z=4.90$ quasar SDSS~0211$-$0009. 
We have used the observed 0.5--2 keV flux of 3.1 $\times$ 10$^{-15}$ \hbox{erg cm$^{-2}$ s$^{-1}$} and 
assumed a $\Gamma=2$ power-law spectrum with Galactic absorption. 
A narrow neutral \hbox{iron K$\alpha$} line has been 
included with a rest-frame equivalent width of 140 eV. 
\label{fig7}}
\centerline{}
\centerline{}
\centerline{}
\noindent
will probably require 
future missions such as \CONS and especially {\it XEUS}. 
Consider, for example, the typical RQQ SDSS~0211$-$0009 at $z=4.90$. 
Given our results on this quasar in Tables~2 and 3, \XMM would gather $\approx260$ EPIC PN$+$MOS 
counts (0.2--10 keV) in a \hbox{40 ks} observation; 
these would contrain its basic spectral properties, but spectral fitting would be limited. 
For comparison, 40 ks \CONS and \XEUS observations would gather $\approx860$ \hbox{(0.5--10 keV)} 
and $\approx31000$ (0.2--10 keV) counts, respectively. 
In Fig.~7 we show simulated \hbox{40 ks} spectra of SDSS~0211$-$0009 from 
\XEUS (with the Cryogenic Imaging Spectrometers, final configuration) and \CONS (with the calorimeter). 
\XEUS and \CONS will be able to measure the X-ray power-law photon
index with an accuracy of $\approx 1$\% and $\approx 8$\%, respectively 
(for comparison, \XMM can measure the photon index to $\approx 25$\%). 
Such measurements will allow quantitative investigation of the 
postulated evolution of the X-ray continua of quasars with redshift; 
at present, such studies are limited to objects at $z\simlt 2.5$ where
the power-law photon index may flatten by $\Delta\Gamma\approx 0.2$ 
between $z\approx 0$ and $z\approx 2.5$ (Vignali et~al. 1999).
\XEUS can detect absorption by neutral material with a column density 
larger than $10^{21}$~cm$^{-2}$ at the source redshift, so it should 
be able to probe clearly any evolution in the amount of absorption 
with redshift. Such absorption measurements will complement the studies 
being made at sub-mm and mm wavelengths, hopefully revealing the nascent 
material fueling black holes in their growth phase. 
\XEUS will also be able to study the inner part of the accretion 
flow via the broad iron~K$\alpha$ line. Given a 6.4~keV iron~K$\alpha$ 
line with a rest-frame equivalent width of 200~eV and a width 
of $\sigma=$0.43~keV (the average width obtained by Nandra et al. 1997 for 
a sample of Seyfert~1 galaxies), \XEUS can measure the line energy, equivalent 
width, and width with an accuracy of $\approx 2$\%, $\approx 29$\% 
and $\approx 29$\%, respectively. 
\vglue 0.3cm

\section{Summary}

We have presented the first results from an exploratory \CHANDRA program 
aimed to measure the X-ray properties of the highest-redshift quasars. 
Ten objects were successfully detected, while four 
were not detected. 
Four of the five highest-redshift X-ray detections to date are presented in 
this work, and the number of $z>4$ X-ray detected quasars has been increased
by 71\%. The only radio-loud quasar in the present sample, 
SDSS~0210$-$0018 ($z=4.77$), has been serendipitously found in one of the 
\CHANDRA fields. To date, it represents the highest-redshift RLQ detected
in the X-ray band. 

Combining the present results with those previously obtained for low-redshift 
samples, we have found that the \CHANDRA quasars are on average X-ray faint 
(as confirmed by the optical-to-X-ray spectral index, $\alpha_{\rm ox}$).
It seems likely that some of the \CHANDRA $z>4$ quasars are surrounded, during 
their growth phase, by large amounts of gas and dust; this matter may both 
feed the newly born massive black holes and weaken the X-ray emission. 
\vglue .5cm

\acknowledgments

We gratefully acknowledge the financial support of \CHANDRA X-ray Center 
grant G01-2100X (CV, WNB, DPS), the Alfred P. Sloan Foundation (WNB), 
NSF grant PHY00-70928 and a Frank and Peggy Taplin Fellowship (XF), 
NASA LTSA grant NAG5-8107 (SK), NSF grant AST99-00703 (DPS), and 
NSF AST00-71091 (MAS). 
CV also acknowledges partial support from the Italian Space Agency, under the contract 
ASI 00/IR/103/AS, and from the Ministry for University and Research (MURST) 
under grant Cofin-00-02-36. 
We thank D. Alexander for help with IDL codes, 
P. Green for providing us with LBQS data, L. Townsley for the CTI correction, 
E. Feigelson and K. Weaver for useful discussions, 
and an anonymous referee for useful comments. 

The Hobby-Eberly Telescope (HET) is a joint project of the University of Texas 
at Austin, the Pennsylvania State University, Stanford University, 
Ludwig-Maximillians-Universit\"at M\"unchen, and Georg-August-Universit\"at G\"ottingen. 
The HET is named in honor of its principal benefactors, 
William P. Hobby and Robert E. Eberly. 


\end{document}